\documentclass[prb, twocolumn, superscriptaddress, showpacs,amsmath,amssymb]{revtex4}

\usepackage{float}
\usepackage{graphicx}
\usepackage{subfigure}
\usepackage{verbatim}

\begin{document}

\title{Assignment of the NV$^0$ 575 nm zero-phonon line in diamond to a $^2$E  -  $^2$A$_2$ transition}
\author{N. B. Manson}
\affiliation{Laser Physics Center, RSPE, Australian National University, Canberra, ACT 0200, Australia}
\author{K. Beha}
\author{A. Batalov}
\affiliation{Department of Physics and Center for Applied Photonics, University of Konstanz, D-78457 Konstanz, Germany}
\author{L. J. Rogers}
\author{M. W. Doherty}
\affiliation{Laser Physics Center, RSPE, Australian National University, Canberra, ACT 0200, Australia}
\author{R. Bratschitsch}
\affiliation{Institute of Physics, Chemnitz University of Technology, D-09107 Chemnitz, Germany}
\author{A. Leitenstorfer}
\affiliation{Department of Physics and Center for Applied Photonics, University of Konstanz, D-78457 Konstanz, Germany}

\email{Neil.Manson@anu.edu.au}

\begin{abstract}
The time-averaged emission spectrum of single nitrogen-vacancy defects in diamond gives zero-phonon lines of both the negative charge state at 637 nm (1.945 eV) and the neutral charge state at 575 nm (2.156 eV). This occurs through photo-conversion between the two charge states. Due to strain in the diamond the zero-phonon lines are split and it is found that the splitting and polarization of the two zero-phonon lines are the same. From this observation and consideration of the electronic structure of the nitrogen-vacancy center it is concluded that the excited state of the neutral center has A$_2$ orbital symmetry. The assignment of the 575 nm transition to a $^2$E - $^2$A$_2$ transition has not been established previously.
\end{abstract}

\maketitle

The nitrogen-vacancy (NV) center has attracted much attention due to many applications as a single photon source,\cite{Kurtsiefer,Aharonovich} in quantum information processing,\cite{Neumann,Togan,Robledo} in magnetometry\cite{Balasubramanian,Maze} and in bio-labelling.\cite{Han,Tisler,McGuiness} The center can exist in a neutral (NV$^0$) or a negative (NV$^-$) charge state and there is interest in the electronic energy levels of both of these charge states. In the case of NV$^-$, the levels and their symmetries are well established. However, the situation for NV$^0$ is not so well advanced. Indeed, the orbital symmetry of the excited state, which gives rise to an orange emission with a zero-phonon line (ZPL) at 575 nm (2.156 eV),  has not been conclusively established. In this work, the symmetry of this level is determined by the analysis of the stress splitting of the NV$^0$ ZPL. The one-electron orbitals of the NV center are well known from the work on NV$^-$ and here they are used to develop the multi-electron orbitals of NV$^0$ in order to account for the observed correlation between the stress splittings of the NV$^0$ and NV$^-$ optical ZPLs.

The NV center is formed in diamond containing singly substitutional nitrogen (N$_{s}$) by irradiating and annealing. \cite{Davies76} The irradiation must have sufficient energy to create vacancies. The vacancies are mobile at temperatures above 600$^\circ$C and become trapped by the nitrogen impurities to form NV pairs. There are five non-bonded electrons when the pair is in the neutral charge state; two from the nitrogen and three from the adjacent carbons. When there are donors in the diamond an extra electron may be acquired to form the negatively charged center with six electrons. Both neutral and negatively charged NV centers have a single allowed transition in the visible and the presence of these defect centers is usually established by the observation of a ZPL at 575 nm in the case of NV$^{0}$ and at 637 nm in the case of NV$^{-}$. Both lines are accompanied by vibrational sidebands of the order of 100 nm to higher energy in absorption and lower energy in emission.\cite{Davies76,Davies79} Excitation at the wavelength of the ZPL or the sideband can convert the center via a quadratic process to the alternate charge state. This is a well established phenomenon for low nitrogen doped samples. \cite{Beha,Aslam} When exciting a single center at a wavelength that coincides with the sideband of both of the charge states the one center is converted back and forth between the two charge states and both centers are observed in a time-average spectrum.\cite{Beha} This is the situation here.

\begin{figure}
\includegraphics[width=0.9\columnwidth]{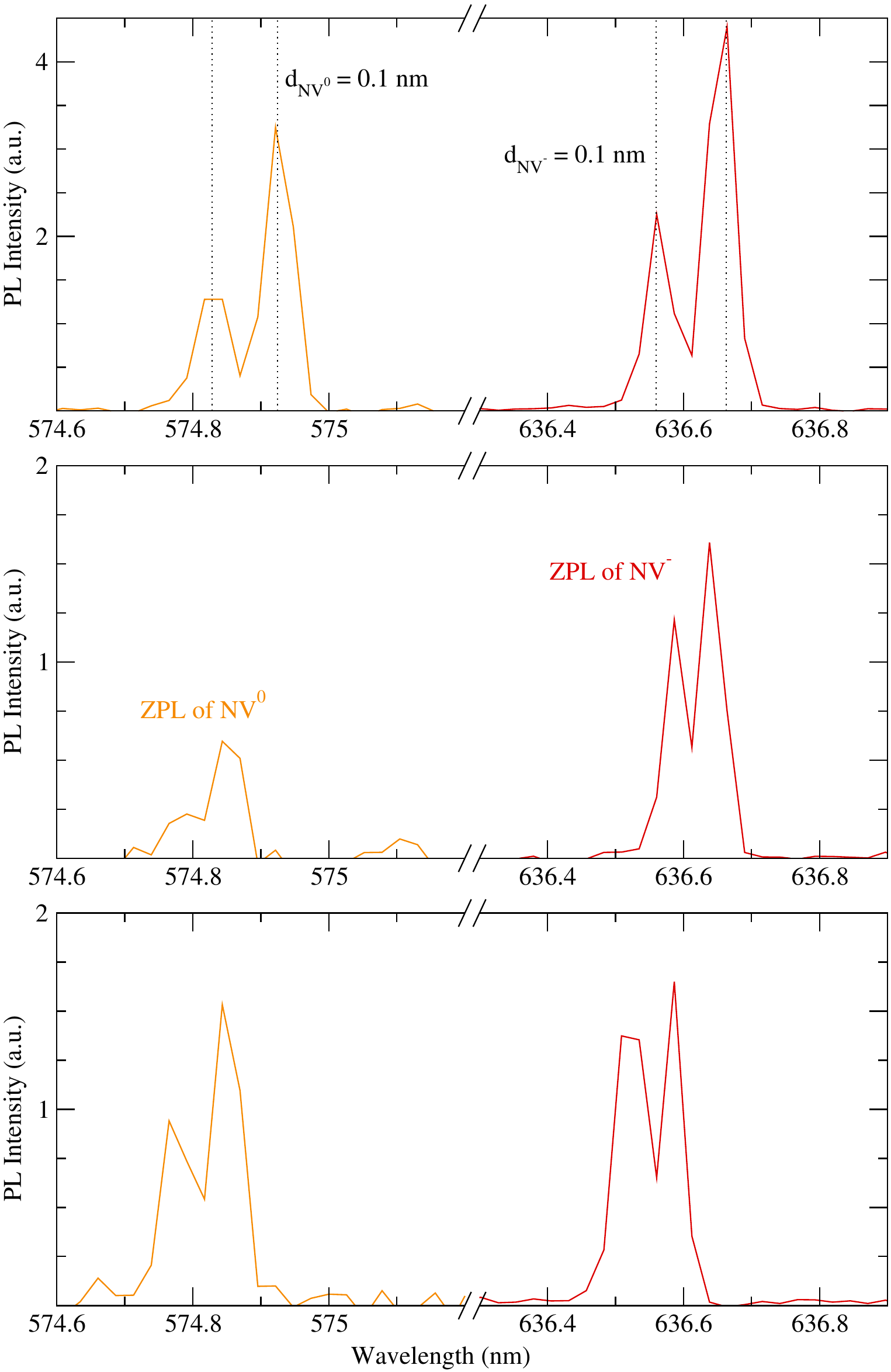}
\caption{Emission spectra of single NV centers averaged over 100 s. In all cases there are ZPLs associated with NV$^-$ at $~$ 635.6 nm (red) and NV$^0$ at $~$ 574.9 nm (orange). For a single site the splittings of the NV$^-$  and NV$^0$ ZPLs are the same although the magnitude can vary from site to site.}
\label{figure 4}
\end{figure}

In the present experiment the emission of a low nitrogen doped sample is monitored during continuous excitation at 532 nm, which is a wavelength within the vibrational absorption sideband of both NV$^-$ and NV$^0$. The sample is ultra pure diamond with nitrogen concentration below 5 ppb. The detection system  involves a low temperature (10 K) confocal microscope system described previously.\cite{Beha} The emission was detected in the range from 550 nm to 650 nm and it can be seen from Fig. 1 that when averaging over 100 s both ZPLs at $\sim$637 nm and $\sim$575 nm, associated with  NV$^-$ and NV$^0$, respectively, are detected. Autocorrelation measurements were used to ensure the emission was associated with a single site. The samples exhibit strain and this results in both of the ZPLs being split. As the emission lines are associated with a single center the splittings of the $\sim$575 nm and $\sim$637 nm ZPLs are clearly associated with identical strains. The splittings vary from site to site but for a single center the splittings of the two ZPLs are the same within the experimental resolution of 0.01 nm. This is illustrated for three different centers in Fig. 1, where the splittings vary but are of similar magnitude $\sim$ 0.1 nm (0.3 meV). There is always correspondence in the magnitude of the ZPL splittings of NV$^-$ and NV$^0$ and this indicates that the effect of internal strain in the diamond is larger than that from any modification at the site due to the change of charge state.

\begin{figure}
\includegraphics[width=0.9\columnwidth]{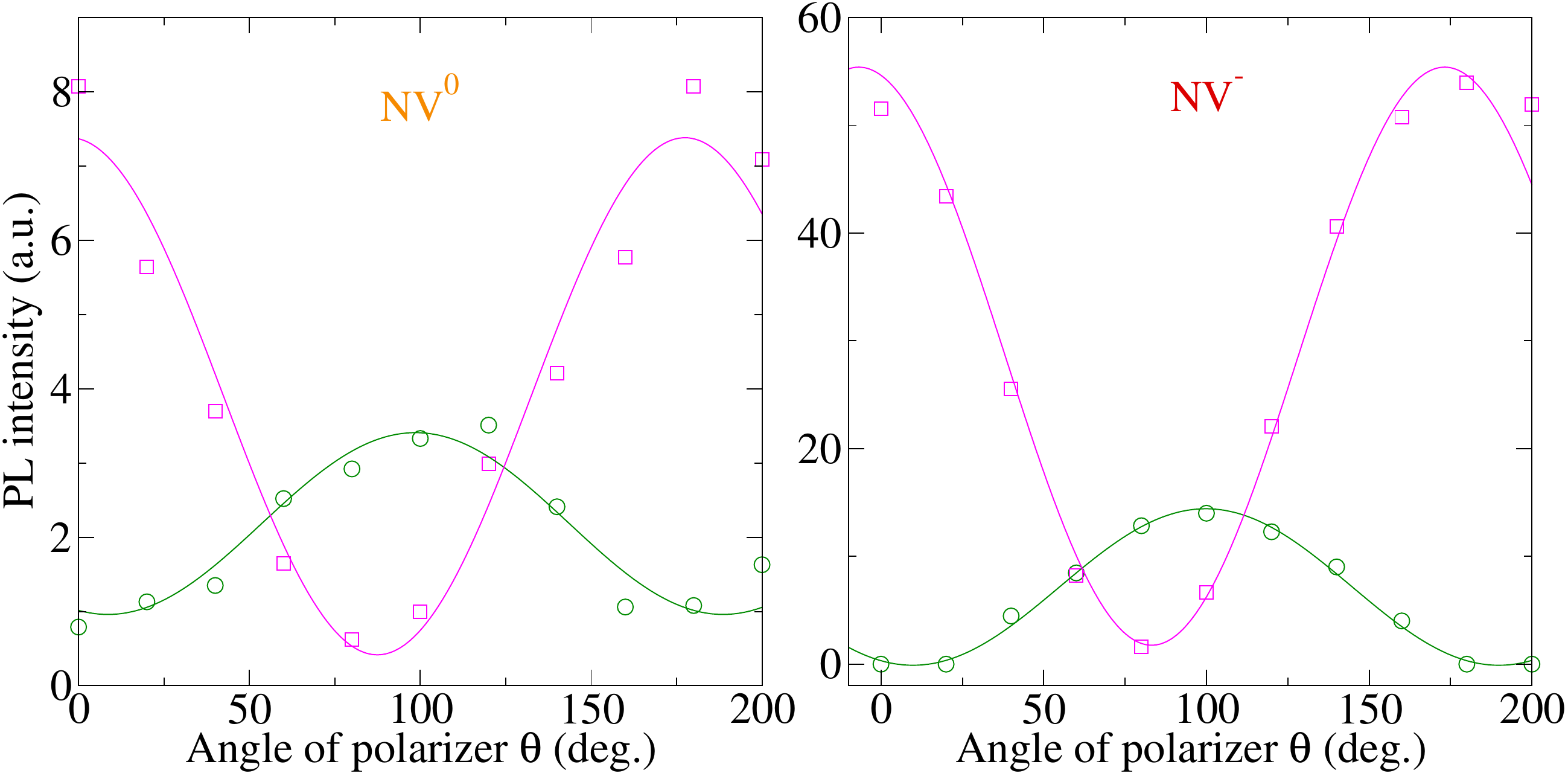}
\caption{Polarization of NV ZPLs for single center shown in the upper figure of Fig. 1. Angle of polarizer for NV$^0$ is on left hand figure and that of NV$^-$ on the right. Experimental data is given in magenta (high energy line) and green (low energy line). The lines are cosine fits to the data.}
\label{figure}
\end{figure}

The polarization of the zero-phonon emission was also measured. This involved using a polarizer to rotate the polarization direction of the detected emission. The result of a systematic rotation of the detected polarization direction is shown in Fig. 2. As indicated by the figure the lines are linearly polarized. The high energy ZPL components for each charge state have one polarization and the polarization of the low energy components are shifted  by 74 $\pm$ 2 degrees. It is well known that for a transition between non-degenerate and degenerate levels at a site with axial symmetry, that transverse strain can lift the degeneracy and give two transitions with polarization directions that are parallel and perpendicular to the direction of the transverse strain. In the present geometry, the photoluminescence detection, and thus the polarization measurements, are for emission perpendicular to a $\langle$100$\rangle$ face, whereas the NV centers are trigonally aligned along $\langle$111$\rangle$ directions. An angle of observation inclined from the NV axis will introduce an apparent phase shift between the orthogonal polarizations of the two transitions. The direction of the local strain of the NV center can be determined from the phase shift. Indeed, Grazioso et al.\cite{Grazioso} have shown by measuring a number of adjacent NV$^-$ centers (including ones with different orientations) that it is possible to determine the local strain field of the diamond lattice. This is not the purpose here. The significant observations are that all sites exhibit the ZPLs of both NV$^0$ and NV$^-$ and that the ZPLs of both charge states are nearly identically split and have the same emission polarization.

\begin{figure}
\includegraphics[width=0.9\columnwidth]{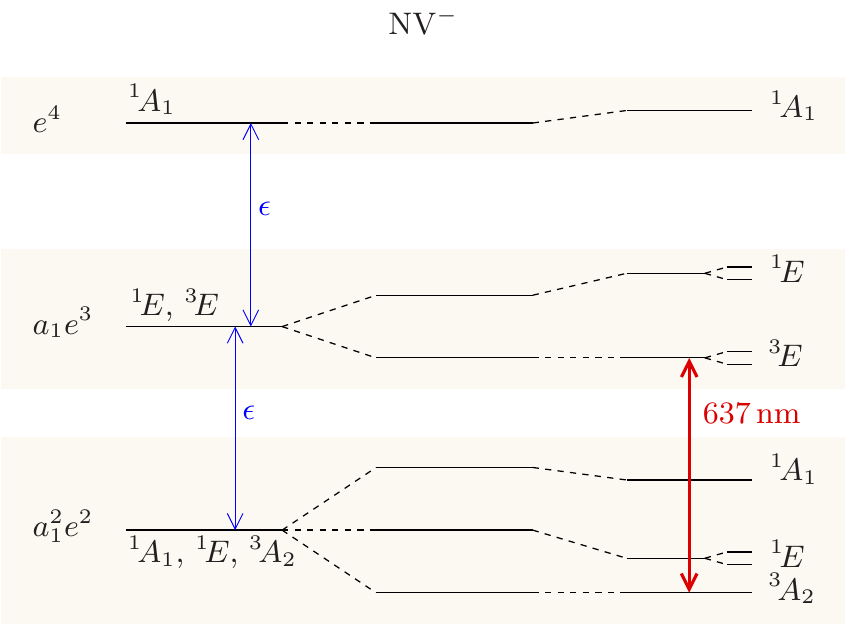}
\caption{Energy levels of the lowest energy configurations of NV$^-$. The diagram depicts configuration levels (left hand side) and first order (central) and second order (right hand side) corrections due to two-electron Coulomb repulsion. Splitting of the orbital E states that can occur with strain are also indicated. The vertical arrow on the left hand side denotes the difference in energy $\epsilon$ of the $e$ and $a_1$ one-electron orbitals. The vertical arrow on the right indicates the spin-allowed optical transition from the ground state. The configurations are given in terms of four electrons. The two additional electrons in the valence band can be ignored.}
\label{figure}
\end{figure}

The electronic structure of the NV center has been modelled by considering the one-electron symmetry adapted orbitals associated with the nitrogen and carbon atoms adjacent to the vacancy. The center's point symmetry is C$_{3v}$ and there are four one-electron orbitals, two degenerate orbitals of E symmetry denoted by e$_x$ and e$_y$ and two orbitals of A$_1$ symmetry denoted by a$_1$ and a$_1^\prime$. The a$_1^\prime$ orbital lies in the valence band, is always occupied and can be neglected. Only the occupation of the a$_1$ and e orbitals in the band gap need be considered. The a$_1$ lies lower and separated from the e orbital by an energy $\epsilon$.
In the case of NV$^-$ there are four electrons occupying the a$_1$ and e orbitals and, in order of increasing energy, they form the configurations a$_1^2$e$^2$ ($^3$A$_2$, $^1$E, $^1$A$_1$), a$_1$e$^3$ ($^3$E, $^1$E) and e$^4$ ($^1$A$_1$), which are to zero order in two-electron Coulomb repulsion are separated by increments of $\epsilon$. The multi-electron levels of the NV$^-$ center are then readily determined and have been given in brackets. Considering first order Coulomb repulsion corrections, Hund's rule gives the triplet to be the lowest energy level within the two lower energy the configurations. Second order Coulomb repulsion corrections shift only the two $^1$E and $^1A_1$ levels. The energy levels are then as indicated in Fig. 3.\cite{Doherty,Maze11} The electric dipole operator conserves spin and as the ground state has a spin S=1 and there is only the one excited triplet level, the electric dipole allowed optical transition giving the ZPL at  $\sim$637 nm  can be assigned to the $^3$A$_2$(a$_1^2$e$^2$)  -   $^3$E(a$_1$e$^3$) transition.

\begin{figure}
\includegraphics[width=0.9\columnwidth]{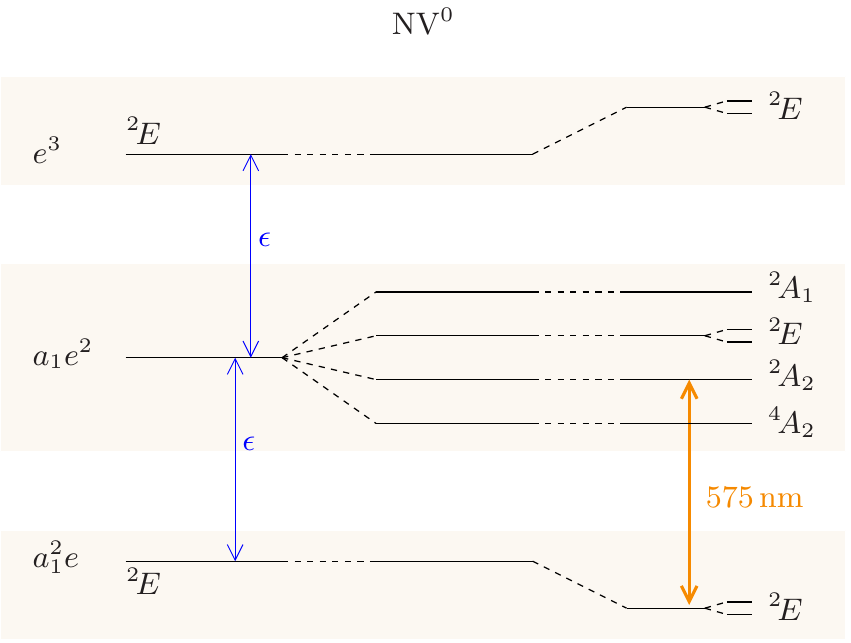}
\caption{Energy level of the lowest energy configuration of NV$^0$. The diagram depicts configuration levels (left hand side) and first order (central) and second order (right hand side) corrections due to two-electron Coulomb repulsion. Splitting of the orbital E states that can occur with strain are also indicated. The vertical arrow on the left hand side denotes the difference in energy $\epsilon$ of the $e$ and $a_1$ one-electron orbitals. The vertical arrow on the right indicates the spin-allowed optical transition from the ground state. The configurations are given in terms of three electrons. The two additional electrons in the valence band can be ignored. Note that the ordering of the $^2$A$_2$, $^2$E and $^2$A$_1$ levels of the a$_1$e$^2$ configuration are unknown without experimental investigation. }
\label{figure}
\end{figure}

For NV$^0$ three electrons occupy the one-electron orbitals and, in order of increasing energy, form the configurations (and levels)  a$_1^2$e ($^2$E), a$_1$e$^2$ ($^4$A$_2$, $^2$A$_2$, $^2$E and $^2$A$_1$) and e$^3$ (E$^2$) (refer to Fig. 4).  Considering first order Coulomb repulsion corrections, within the intermediate configuration, Hund's rule determines that the quartet will have lower energy than the three doublets, but provides no information about the ordering of the doublets. There are also second order Coulomb repulsion corrections between all of the $^2$E levels. As a consequence, the energy ordering of the doublets of the a$_1$e$^2$ configuration can not be readily determined. Determining their order is significant as the emitting level is likely to be the lowest state with the same spin as the ground state. The optical transition from the ground state to this level gives the ZPL at $\sim$575 nm and from uniaxial stress measurements, this ZPL has been shown to correspond to a transition from an E ground state to an A excited state.\cite{Davies79}  The uniaxial stress measurements, however, cannot determine whether the non-degenerate state has A$_1$ or A$_2$ orbital symmetry.

The NV$^-$ ZPL at $\sim$637 nm is, therefore, definitely associated with an $^3$A$_2$(a$_1^2$e$^2$) - $^3$E(a$_1$e$^3$) transition whereas the NV$^0$ ZPL at $\sim$575 nm may be associated with either a $^2$E(a$_1^2$e) - $^2$A$_1$(a$_1$e$^2$) or $^2$E(a$_1^2$e) - $^2$A$_2$(a$_1$e$^2$) transition. As there can be no splitting of the non-degenerate A levels, the strain splittings of the ZPLs must arise from splittings of the degenerate E levels. \cite{Davies76,Davies79} From the electronic model of the NV center, the splitting of the multi-electron levels will arise from the splitting of the one-electron e orbitals. In the case of NV$^0$, the splitting is in the lower $^2$E(a$_1^2$e) level, where a single electron occupies the e orbitals, whereas for NV$^-$, the splitting is in the excited $^3$E(a$_1$e$^3$) level, where three electrons occupy the e orbitals. As four e orbital electrons (e$^4$) is a full shell, the latter may be considered as an e orbital hole. The situation can therefore then be summarised as in Fig. 5, with splitting of an e-electron in the $^2$E ground state of NV$^0$ and e-hole in the excited $^3$E level of NV$^-$. Clearly the splitting of a hole will be the reverse of, but equal to, that of an electron due to the charges being opposite in sign. This is analogous to the change of sign of the Zeeman splitting between an electron and a hole. The change of sign is indicated in Fig. 5 and further justified below.

For a strain along a direction X (perpendicular to center axis Z but not necessarily a symmetry direction), the one-electron e$_X$ orbital is shifted up in energy and the e$_Y$ orbital is shifted down, resulting in a splitting of the e orbitals. Using the conclusions of the previous paragraph, this implies that the NV$^0$ ground $^2$E (equivalent to an e-electron) level is split such that the $^2$E$_X$ state is shifted up and the $^2$E$_Y$ state is shifted down. The converse is true for the excited $^3$E level of NV$^-$ (equivalent to an e-hole), where the $^3$E$_Y$ state is shifted up and the $^3$E$_X$ state is shifted down. Considering $C_{3v}$ symmetry selection rules for an electric dipole transition, an A$_2$ - E$_X$ transition is Y polarized (orthogonal to X and Z) and an A$_2$ - E$_Y$ transition is X polarized.\cite{Davies76,Davies79} These selection rules define the polarizations of the transitions and are depicted in Fig. 5. Since the splittings of the $^2$E (NV$^0$) and $^3$E (NV$^-$) levels are equal and opposite, but they occur in ground and excited levels of the optical transitions, respectively, it is evident that the ZPLs of NV$^0$ and NV$^-$ will be split by approximately the same energy and that the higher energy components of the NV$^0$ and NV$^-$ ZPLs will be both $X$ polarized and the lower energy components will be both $Y$ polarized. This result matches experimental observation. Should the excited state of NV$^0$ be $^2$A$_1$, as sometimes suggested, the polarization of the high and low energy components of the NV$^0$ and NV$^-$ ZPL would each differ. This is not the case and clearly the observations prove that the emitting level of the NV$^0$ has A$_2$ orbital symmetry.

\begin{figure}
\includegraphics[width=0.9\columnwidth]{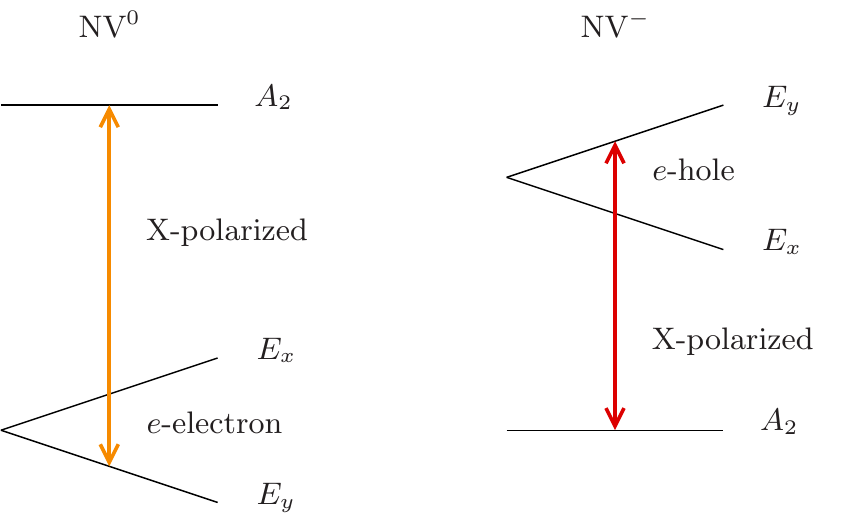}
\caption{Summary of splitting due to stress for the $\sim$575 nm NV$^0$ and $\sim$637 nm NV$^-$ transitions. A transition with a common X-polarization is shown and in both cases is displaced to higher energy. The transition to the E$_X$ state is Y-polarized and displaced to lower energy but, for clarity, is not shown.}
\label{figure 2}
\end{figure}

Stress measurements by themselves cannot distinguish between A$_1$ - E and A$_2$ - E transitions and this has been made very clear  by Davies in his discussion of the results of uniaxial studies of NV$^-$ \cite{Davies76} and NV$^0$.\cite{Davies79} In the case of NV$^-$ the uniaxial stress indicated that the ground state had  orbital A symmetry \cite{Davies76} but the $^3$A$_2$ assignment can only be made once the ground state has been identified to have  S =1 and modelling indicated that there is only one spin triplet in the ground configuration. It is a similar situation here for NV$^0$, in that the $^2$A$_2$ assignment is required to give consistency between the electronic modelling of NV$^0$ and the model that is now established for NV$^-$.

The latter A$_2$ assignment could in principle be deduced from the uniaxial stress studies of the $\sim$637 nm \cite{Davies76} and $\sim$575 nm \cite{Davies79} ZPLs involving ensembles and applied stress of several GPa. Certainly the magnitudes of the splitting parameters as a function of stress are better determined using controlled external stress and can provide extra information about the shift of levels involving both initial and final states. The importance of using single site measurements is that there can be no uncertainty about the strain conditions and unquestionably the observed splitting and polarization are for identical strain fields. The equivalence of the splittings for the NV$^-$ and NV$^0$ ZPLs in the single site measurements such as in Fig. 1 is striking.

The lowest excited doublet and emitting level of NV$^0$ is frequently discussed in terms of being an $^2$A$_1$ rather than $^2$A$_2$ and often this is of little consequence.  However, the misidentification does give difficulty in accounting for the population of the low lying quartet $^4$A$_2$ observed in electron paramagnetic resonance when the NV$^0$ center is optically excited.\cite{Felton} There is strong evidence that the quartet is populated from the emitting level at 2.156 eV, but there is no spin-orbit mixing between the $^2$A$_2$ and $^4$A$_2$ levels (as they both have A$_2$ orbital symmetry) and so the decay of population from $^2$A$_2$ to $^4$A$_2$ cannot arise from a conventional inter-system crossing. Clearly it is important to allow for a $^2$A$_2$ level when calculating energy levels \cite{Gali09,Ranjbar} and it is noted that the presence of an $^2$A$_2$ can be overlooked. There is only the one calculation of Zyubin et al.\cite{Zyubin} that indicates the lowest energy excited doublet is an $^2$A$_2$. When allowing for the vibronic band their energy of 2.4 eV is in good correspondence with the experiment. The correspondence in the magnitude of the splittings observed for NV$^0$ and NV$^-$  is surprising since the charge states have different numbers of electrons and therefore different electron configurations. Additionally, some redistribution of the local environment of the center is expcted to occur in the photoconversion process between charge states. However, the observed correlations of the ZPLs of the charge states provides an ideal situation for comparison with theoretical calculations, as demonstrated here.

In short, the correlated energy splittings and polarizations of the NV$^-$ and NV$^0$ optical ZPLs observed in the spectrum of a single NV color center in diamond is quite striking and, as shown here, imply that both transitions occur between levels of A$_2$ and E orbital symmetry. This leads to the first firm assignment of the $^2$A$_2$  level to the lowest energy doublet and optically emitting level of NV$^0$.

\begin{acknowledgments}
The work was partly supported by Australian Research Council grant DP12010223. Support by the Deutsche Forschungsgemeinschaft (FOR 1493) is gratefully acknowledged.
\end{acknowledgments}

\end{document}